\documentstyle[preprint,prl,aps,epsfig]{revtex}
\narrowtext
\draft
\begin{document}

\title{LDA++ approach to electronic structure of magnets:\\
 correlation effects in iron.}

\author{
M. I. Katsnelson$^1$ and A.\ I. Lichtenstein$^2$
}

\address{
$^1$ Institute of Metal Physics, 620219 Ekaterinburg , Russia \\
$^2$ Forschungszentrum J\"ulich, 52425 J\"ulich, Germany
}
\maketitle
\begin{abstract}
A novel approach to investigation of correlation effects in the electronic
structure of magnetic crystals which takes into account a frequency
dependence of the self energy (so called ``LDA++ approach'') is developed.
The fluctuation exchange approximation is generalized to the spin-polarized
multi-band case and its local version is proposed. As an example, we
calculate the electronic quasiparticle spectrum of ferromagnetic iron. It
is shown that the Fermi liquid description of the bands near the Fermi level is
reasonable, while the quasiparticle states beyond approximately 1 eV range
are strongly damped, in agreement with photoemission data. The result of
the spin-polarized thermoemission experiment is explained satisfactory.
The problem of satellite structure is discussed.
\end{abstract}

\section{Introduction}

The description of correlation effects in electronic structure and magnetism
of iron-group metals is still far from the final picture and attracts continuous
interest (see, e.g., \cite{liebcsh,gyorffy,steiner,IKT} and Refs therein). Despite
many attempts, the situation is still unclear both theoretically and
experimentally. For example, there is no agreement on the presence of 5 eV
satellite in photoemission spectrum of iron \cite{satyes,kirby}, and on the
existence of local spin splitting above Curie temperature of nickel \cite
{splitni}. The experimental data on the absence of spin-polarization in the
thermoemission from cesiated iron \cite{thermoem} are still not understood
completely \cite{monnier}. From the theoretical point of view, different
approaches such as the second-order perturbation theory \cite{treglia,steiner},
the three-body Faddeev approximation \cite{manghi},
and the  moment expansion method \cite{nolting} were used.
Unfortunately, the conditions of applicability of these
schemes are not clear. Recently we investigated different approximations to
the LDA-based correlated electronic structure of crystals 
with a local self-energy \cite{lda++} 
(so called ``LDA++'' approach) and  argued that for
moderately strong correlations (the case of iron-group metals) one of the most
efficient approaches would be the fluctuation-exchange (FLEX) approximation of
Bickers and Scalapino \cite{FLEX}. Here we generalize LDA++ approach \cite
{lda++} to the spin-polarized case and present some results for the
quasiparticle spectrum of ferromagnetic iron.

\section{Spin-polarized multi-band FLEX approximation}

Let us start with the general many-body Hamiltonian for crystal in LDA+U
scheme\cite{revU}:

\begin{eqnarray}
H &=&H_t+H_U  \label{ham} \\
H_t &=&\sum\limits_{\lambda \lambda ^{\prime }\sigma }t_{\lambda \lambda
^{\prime }}c_{\lambda \sigma }^{+}c_{\lambda ^{\prime }\sigma }  \nonumber \\
H_U &=&\frac 12\sum\limits_{\left\{ \lambda _i\right\} \sigma \sigma
^{\prime }}\left\langle \lambda _1\lambda _2\left| v\right| \lambda
_1^{\prime }\lambda _2^{\prime }\right\rangle c_{\lambda _1\sigma
}^{+}c_{\lambda _2\sigma ^{\prime }}^{+}c_{\lambda _2^{\prime }\sigma
^{\prime }}c_{\lambda _1^{\prime }\sigma \, ,  \nonumber }
\end{eqnarray}
where $\lambda =im$ are the site number $\left( i\right) $ and orbital $%
\left( m\right) $ quantum numbers; $\sigma =\uparrow ,\downarrow $ is the
spin projection; $c^{+},c$ are the Fermi creation and annihilation
operators; $H_t$ is the  effective single particle Hamiltonian from LDA,
corrected for double-counting of average interactions among correlated
electrons\cite{revU,lda++}, and the Coulomb matrix elements
are defined in the standard way

\begin{equation}
\left\langle 12\left| v\right| 34\right\rangle =\int d{\bf r}d{\bf r}%
^{\prime }\psi _1^{*}({\bf r})\psi _2^{*}({\bf r}^{\prime })v\left( 
{\bf{r-r}}^{\prime }\right) \psi _3({\bf r})\psi _4({\bf r}^{\prime
}),  \label{coulomb}
\end{equation}
where we define for briefness $\lambda _1\equiv 1$ etc. 
Following Bickers and Scalapino \cite{FLEX} we introduce the
pairwise operators corresponding to different channels, namely,
particle-hole density:

\[
d_{12}=\frac 1{\sqrt{2}}\left( c_{1\uparrow }^{+}c_{2\uparrow
}+c_{1\downarrow }^{+}c_{2\downarrow }\right) ,
\]
particle-hole magnetic channel:

\begin{eqnarray*}
m_{12}^0 &=&\frac 1{\sqrt{2}}\left( c_{1\uparrow }^{+}c_{2\uparrow
}-c_{1\downarrow }^{+}c_{2\downarrow }\right)  \\
m_{12}^{+} &=&c_{1\uparrow }^{+}c_{2\downarrow } \\
m_{12}^{-} &=&c_{1\downarrow }^{+}c_{2\uparrow } \, ,
\end{eqnarray*}
particle-particle singlet channel:

\begin{eqnarray*}
s_{12} &=&\frac 1{\sqrt{2}}\left( c_{1\downarrow }c_{2\uparrow
}-c_{1\uparrow }c_{2\downarrow }\right)  \\
\overline{s}_{12} &=&\frac 1{\sqrt{2}}\left( c_{1\uparrow
}^{+}c_{2\downarrow }^{+}-c_{1\downarrow }^{+}c_{2\uparrow }^{+}\right) ,
\end{eqnarray*}
and particle-particle triplet channel:

\begin{eqnarray*}
t_{12}^0 &=&\frac 1{\sqrt{2}}\left( c_{1\downarrow }c_{2\uparrow
}+c_{1\uparrow }c_{2\downarrow }\right)  \\
\overline{t}_{12}^0 &=&\frac 1{\sqrt{2}}\left( c_{1\uparrow
}^{+}c_{2\downarrow }^{+}+c_{1\downarrow }^{+}c_{2\uparrow }^{+}\right)  \\
t_{12}^{\pm } &=&c_{1\uparrow ,\downarrow }c_{2\downarrow ,\uparrow } \\
\overline{t}_{12}^{\pm } &=&c_{1\uparrow ,\downarrow }^{+}c_{2\downarrow
,\uparrow }^{+} \, .
\end{eqnarray*}
\ The bare vertex matrices corresponding to the different channels are defined
as:

\begin{eqnarray}
U_{12,34}^d &=&2\left\langle 13\left| v\right| 24\right\rangle -\left\langle
13\left| v\right| 42\right\rangle  \label{vertex} \\
U_{12,34}^m &=&-\left\langle 13\left| v\right| 42\right\rangle  \nonumber \\
U_{12,34}^s &=&\frac 12\left[ \left\langle 12\left| v\right| 34\right\rangle
+\left\langle 12\left| v\right| 43\right\rangle \right]  \nonumber \\
U_{12,34}^t &=&\frac 12\left[ \left\langle 12\left| v\right| 34\right\rangle
-\left\langle 12\left| v\right| 43\right\rangle \right] .  \nonumber
\end{eqnarray}

Now we can rewrite the interaction part of Hamiltonian in Eq.(\ref{ham}) in
two equivalent forms, corresponding to the particle-hole and 
the particle-particle channels:

\begin{eqnarray*}
H_U &=&\frac 12Tr\left\{ d*U^d*d+\sum_{\alpha =0,\pm }m^\alpha
*U^m*m^{-\alpha }\right\}  \\
H_U &=&\frac 12Tr\left\{ \overline{s}*U^s*s+\sum_{\alpha =0,\pm }\overline{t}%
^\alpha *U^t*t^\alpha \right\}, 
\end{eqnarray*}
where ($*$) means matrix product, e.g.

\begin{eqnarray*}
\left( U^d*d\right) _{12} &=&\sum_{34}U_{12,34}^dd_{34} \\
\left( d*U^d\right) _{12} &=&\sum_{34}d_{34}U_{34,12}^d \,.
\end{eqnarray*}

Then we may repeat the usual derivation of FLEX equations for single band \cite
{FLEX} and multi-band \cite{lda++,mbick} cases, taking into account
the  spin-dependence of the Green function
$G_{\lambda \lambda ^{\prime }}^\sigma$:

\[
G_{12}^\sigma \left( \tau \right) =-\left\langle T_\tau c_{1\sigma }\left(
\tau \right) c_{2\sigma }^{+} \left(0 \right)\right\rangle .
\]

For the  finite temperature (T$>0$) FLEX equations has a ``local form''
in the Matsubara frequency ($i\omega _n $) or imaginary time ($\tau $) space
(wherwheree $\omega _n=(2n+1)\pi T,$ $n=0,\pm 1,...$) and it is  
very efficient to use the fast-Fourier transforms (FFT) with periodic boundary
condition\cite{FLEX,lda++}. Time-frequency spaces are connected by

\begin{eqnarray*}
G_{\lambda \lambda ^{\prime }}^\sigma
(i\omega _n) &=&\int_0^\frac 1T e^{i\omega _n\tau }
G_{\lambda \lambda ^{\prime }}^\sigma (\tau )d\tau \\
G_{\lambda \lambda ^{\prime }}^\sigma
(\tau ) &=& T \sum_{\omega _n}e^{-i\omega _n\tau }
G_{\lambda \lambda ^{\prime }}^\sigma
(i\omega _n) \, ,
\end{eqnarray*}
and we will try to keep this dual $(i\omega) -(\tau )$ notation to stress the
numerical implementation of this LDA++ scheme. 

An unusual feature of the spin-polarized multi-band FLEX scheme 
is mixing of $m^0$ and $d$-channels (as well as $s$
and $t^0$-channels) which related with non-zero value
of such correlator, for example: 
\begin{eqnarray}
\left\langle \left\langle d_{12}m_{34}^0\right\rangle \right\rangle _c
=\frac 12\left\langle \left\langle \left( c_{1\uparrow }^{+}c_{2\uparrow
}+c_{1\downarrow }^{+}c_{2\downarrow }\right) \left( c_{3\uparrow
}^{+}c_{4\uparrow }-c_{3\downarrow }^{+}c_{4\downarrow }\right)
\right\rangle \right\rangle _c=  \label{connect} 
\ -\frac 12\left\{ G_{23}^{\uparrow }G_{41}^{\uparrow }-G_{23}^{\downarrow
}G_{41}^{\downarrow }\right\} , 
\end{eqnarray}
where ``$c$'' denotes a connected part of the correlator (cf. \cite{FLEX}). As
a result, we have the following expression for effective
transverse susceptibility matrix:

\begin{equation}
\chi ^{+-}(i\omega )=\left[ 1+U^m*\Gamma ^{\uparrow \downarrow }(i\omega
)\right] ^{-1}*\Gamma ^{\uparrow \downarrow }(i\omega ) \, , \label{xi+-}
\end{equation} 
where

\begin{equation}
\Gamma _{12,34}^{\sigma \sigma ^{\prime }}\left( \tau \right)
=-G_{23}^\sigma \left( \tau \right) G_{41}^{\sigma ^{\prime }}\left( -\tau
\right)  \label{gamma}
\end{equation}
is an ``empty loop'' and $\Gamma (i\omega )$ is its Fourier transform.
The corresponding longitudinal susceptibility matrix has a  more complicated form:

\begin{equation}
\chi ^{\parallel }(i\omega )=\left[ 1+U_{ph}^{\parallel }*\chi _0^{\parallel
}(i\omega )\right] ^{-1}*\chi _0^{\parallel }(i\omega ) ,  \label{xipar}
\end{equation}
where we introduce the supermatrix of p-h interactions:

\[
U_{ph}^{\parallel }=\left( 
\begin{array}{cc}
U^d & 0 \\ 
0 & U^m
\end{array}
\right) , 
\]
and the matrix of bare longitudinal susceptibility:

\begin{equation}
\chi _0^{\parallel }=\frac 12\left( 
\begin{array}{cc}
\Gamma ^{\uparrow \uparrow }+\Gamma ^{\downarrow \downarrow }\, & \, \Gamma
^{\uparrow \uparrow }-\Gamma ^{\downarrow \downarrow } \\ 
\Gamma ^{\uparrow \uparrow }-\Gamma ^{\downarrow \downarrow }\, & \, \Gamma
^{\uparrow \uparrow }+\Gamma ^{\downarrow \downarrow }
\end{array}
\right) ,  \label{xi0par}
\end{equation}
in the $dd$-, $dm^0$-, $m^0d$-, and $m^0m^0$- channels ($%
d,m^0=1,2$ in the supermatrix indices). Similarly for the p-p channel we have:

\begin{eqnarray}
R^{\pm}(i\omega ) &=&\left[ 1+U^t*R_0^{\pm }(i\omega )\right] ^{-1}*R_0^{\pm
}(i\omega )  \label{R+-} \\
R^{\parallel }(i\omega ) &=&\left[ 1+U_{pp}^{\parallel }*R_0^{\parallel
}(i\omega )\right] ^{-1}*R_0^{\parallel }(i\omega ) \, , \nonumber
\end{eqnarray}
where the supermatrix of p-p interactions is defined as: 

\[
U_{pp}^{\parallel }=\left( 
\begin{array}{cc}
U^s & 0 \\ 
0 & U^t
\end{array}
\right) , 
\]
and the bare p-p susceptibilities are

\begin{equation}
\left[ R_0^{\pm }\left( \tau \right) \right] _{12,34}=\frac 12\left[
G_{14}^{\uparrow ,\downarrow }\left( \tau \right) G_{23}^{\uparrow
,\downarrow }\left( \tau \right) -G_{13}^{\uparrow ,\downarrow }\left( \tau
\right) G_{24}^{\uparrow ,\downarrow }\left( \tau \right) \right]
\label{R0+-} ,
\end{equation}

\[
R_0^{\parallel }=\left( 
\begin{array}{cc}
R_0^{ss} & R_0^{st} \\ 
R_0^{ts} & R_0^{tt}
\end{array}
\right) , 
\]

\begin{equation}
\Pi _{12,34}^{\sigma \sigma ^{\prime }}\left( \tau \right) =G_{23}^\sigma
\left( \tau \right) G_{14}^{\sigma ^{\prime }}\left( \tau \right) ,
\label{pipp}
\end{equation}

\begin{eqnarray*}
\left[ R_0^{ss}\right] _{12,34} &=&\frac 14\left[ \Pi _{12,34}^{\uparrow
\downarrow }+\Pi _{21,34}^{\uparrow \downarrow }+\Pi _{21,34}^{\downarrow
\uparrow }+\Pi _{12,34}^{\downarrow \uparrow }\right] \\
\left[ R_0^{st}\right] _{12,34} &=&\frac 14\left[ \Pi _{12,34}^{\uparrow
\downarrow }+\Pi _{21,34}^{\uparrow \downarrow }-\Pi _{21,34}^{\downarrow
\uparrow }-\Pi _{12,34}^{\downarrow \uparrow }\right] \\
\left[ R_0^{ts}\right] _{12,34} &=&\frac 14\left[ \Pi _{12,34}^{\uparrow
\downarrow }-\Pi _{21,34}^{\uparrow \downarrow }+\Pi _{21,34}^{\downarrow
\uparrow }-\Pi _{12,34}^{\downarrow \uparrow }\right] \\
\left[ R_0^{tt}\right] _{12,34} &=&\frac 14\left[ \Pi _{12,34}^{\uparrow
\downarrow }-\Pi _{21,34}^{\uparrow \downarrow }-\Pi _{21,34}^{\downarrow
\uparrow }+\Pi _{12,34}^{\downarrow \uparrow }\right] .
\end{eqnarray*}

In the FLEX approximation we can calculate the electronic self-energy in terms of effective
interactions in various channels:

\begin{equation}
\Sigma =\Sigma ^{HF}+\Sigma ^{(2)}+\Sigma ^{(ph)}+\Sigma ^{(pp)} ,
\label{sig}
\end{equation}
where the Hartree-Fock contribution is equal to:

\begin{equation}
\Sigma _{12,\sigma }^{HF}=\sum_{34}\left[ \left\langle 13\left| v\right|
24\right\rangle \sum_{\sigma ^{\prime }}n_{34}^{\sigma ^{\prime
}}-\left\langle 13\left| v\right| 42\right\rangle n_{34}^\sigma \right] ,
\label{shf}
\end{equation}
with the occupation matrix $n_{12}^\sigma=G_{21}^\sigma (\tau \rightarrow -0)$;
this contribution to $\Sigma$ is equivalent to spin-polarized
``rotationally-invariant'' LDA+U method\cite{SPLDA+U}.

The second-order contribution in the spin-polarized case reads:
\begin{eqnarray}
\Sigma _{12,\sigma }^{(2)}(\tau ) =-\sum_{\left\{ 3-8\right\}
}\left\langle 13\left| v\right| 74\right\rangle G_{78}^\sigma (\tau
)\left[ \left\langle 85\left| v\right| 26\right\rangle \sum_{\sigma ^{\prime
}}G_{63}^{\sigma ^{\prime }}(\tau )G_{45}^{\sigma ^{\prime }}(-\tau )
\label{sig2} 
-\left\langle 85\left| v\right| 62\right\rangle G_{63}^\sigma (\tau
)G_{45}^\sigma (-\tau ) \right] , 
\end{eqnarray}
and  the higher-order particle-hole contribution
\begin{equation}
\Sigma _{12,\sigma }^{(ph)}\left( \tau \right) =\sum\limits_{34,\sigma
^{\prime }}W_{13,42}^{\sigma \sigma ^{\prime }}\left( \tau \right)
G_{34}^{\sigma ^{\prime }}\left( \tau \right)   \label{sigph},
\end{equation}
with p-h fluctuation potential matrix:

\begin{equation}
W^{\sigma \sigma ^{\prime }}\left(i\omega \right) =\left[ 
\begin{array}{cc}
W^{\uparrow \uparrow }\left(i\omega \right) & W^{\uparrow \downarrow }\left(i
\omega \right) \\ 
W^{\downarrow \uparrow }\left(i\omega \right) & W^{\downarrow \downarrow
}\left(i\omega \right)
\end{array}
\right] ,  \label{wpp}
\end{equation}
where the spin-dependent effective potentials are defined as

\begin{eqnarray*}
W^{\uparrow \uparrow } &=&\frac 12\{U^d*\left[ \chi ^{dd}-\chi
_0^{dd}\right] *U^d+U^m*\left[ \chi ^{mm}-\chi _0^{mm}\right] *U^m \\
&&+U^d*\left[ \chi ^{dm}-\chi _0^{dm}\right] *U^m+U^m*\left[ \chi ^{md}-\chi
_0^{md}\right] *U^d\}
\end{eqnarray*}

\begin{eqnarray*}
W^{\downarrow \downarrow } &=&\frac 12\{U^d*\left[ \chi ^{dd}-\chi
_0^{dd}\right] *U^d+U^m*\left[ \chi ^{mm}-\chi _0^{mm}\right] *U^m \\
&&\ -U^d*\left[ \chi ^{dm}-\chi _0^{dm}\right] *U^m-U^m*\left[ \chi
^{md}-\chi _0^{md}\right] *U^d\}
\end{eqnarray*}

\[
W^{\uparrow \downarrow }=U^m*\left[ \chi ^{+-}-\chi _0^{+-}\right] *U^m 
\]

\[
W^{\downarrow \uparrow }=U^m*\left[ \chi ^{-+}-\chi _0^{-+}\right] *U^m .
\] 
Finally the higher-order particle-particle contribution corresponds to:
\[
\Sigma _{12,\sigma }^{(pp)}\left( \tau \right) = - \sum\limits_{34,\sigma
^{\prime }}T_{13,42}^{\sigma \sigma ^{\prime }}\left( \tau \right)
G_{43}^{\sigma ^{\prime }}\left(\tau \right) ,
\]
with p-p fluctuation potential matrix:

\begin{equation}
T^{\sigma \sigma ^{\prime }}\left( i\omega \right) =\left[ 
\begin{array}{cc}
T^{\uparrow \uparrow }\left(-i\omega \right) & T^{\uparrow \downarrow }\left(-i
\omega \right) \\ 
T^{\downarrow \uparrow }\left(-i\omega \right) & T^{\downarrow \downarrow
}\left(-i\omega \right)
\end{array}
\right] ,  \label{tpp}
\end{equation}
defined in terms of the spin-dependent p-p effective potentials:

\begin{eqnarray*}
T^{\uparrow \downarrow } &=&U^s*\left[ R^{ss}-R_0^{ss}\right]
*U^s+U^t*\left[ R^{tt}-R_0^{tt}\right] *U^r \\
&&\ +U^s*\left[ R^{st}-R_0^{st}\right] *U^t+U^t*\left[
R^{ts}-R_0^{ts}\right] *U^s ,
\end{eqnarray*}

\begin{eqnarray*}
T^{\downarrow \uparrow } &=&U^s*\left[ R^{ss}-R_0^{ss}\right]
*U^s+U^t*\left[ R^{tt}-R_0^{tt}\right] *U^r \\
&&\ \ -U^s*\left[ R^{st}-R_0^{st}\right] *U^t-U^t*\left[
R^{ts}-R_0^{ts}\right] *U^s ,
\end{eqnarray*}

\[
T^{\uparrow \uparrow }=U^t*\left[ R^{+}-R_0^{+}\right] *U^t ,
\]

\[
T^{\downarrow \downarrow }=U^t*\left[ R^{-}-R_0^{-}\right] *U^t .
\]
Note that for both p-h and p-p channels the effective interactions,
according to Eqs.(\ref{wpp},\ref{tpp}), are non-diagonal matrices in spin
space, in contrast with any mean-field approximation like LSDA.

\section{Local approximation}

The consideration of full non-local FLEX self-energy $\Sigma _{ij}\left(
i\omega \right) $ (or momentum dependent Fourier transform $\Sigma \left( 
{\mathbf k},i\omega \right) $) in the framework of realistic multiband
calculations is faced with significant computational difficulties \cite{mbick}. On
the other hand, the best local approximation which corresponds to the so called
dynamical mean field theory (DMFT) \cite{dinf} appeared to be very effective
for the calculation of electron spectra of strongly correlated systems.
Therefore it is reasonable to combine the FLEX approximation with DMFT rather
than just neglecting momentum dependence in FLEX equations. In this respect
we will take into account only on-site (Hubbard) interactions so all the
bare vertices (Eq. (\ref{vertex})) are diagonal in the site index $i$ but
are matrices in orbital indices $\left\{ m\right\} $.
We believe that this self-consistent local FLEX- approximation
partly accounts for ``vertex corrections'' (due to the
difference of G and $\cal G_0$ see bellow)
 which are absent in usual FLEX
method.
 Moreover, it is known that even simple second-order
approximation for self-energy combined with DMFT give rather good results
for the description of correlated systems namely, Hubbard splitting,
``Kondo resonances'' etc. \cite{dinf}.

The DMFT procedure on the lattice is as follows: one needs to find a
self-consistent solution of functional equations

\begin{eqnarray}
{\mathcal G}_0^{-1}\left( i\omega \right) &=&G^{-1}\left( i\omega \right)
+\Sigma \left( i\omega \right)  \label{dmft} \\
\Sigma &=&\Sigma \left[ {\mathcal G}_0\right]  \, , \nonumber
\end{eqnarray}
where 
\begin{equation}
G\left( i\omega \right) =\sum_{\mathbf{k}} \left[ i\omega +\mu -t\left( 
\mathbf{k}\right) -\Sigma \left( i\omega \right) \right] ^{-1}  \label{dyson}
\end{equation}
is the matrix (in orbital and spin indices) of the local Green function and $%
\Sigma $ (Eq.(\ref{sig})) is our spin-polarized multi-band FLEX solution which 
depends on the effective media Green function ${\cal {G}}_0$.

We could further reduce the computational procedure by neglecting  dynamical
interaction in the p-p channel since the most important fluctuations in
itinerant electron magnets are spin-fluctuations in the p-h channel. We take into
account static ( of $T$- matrix type) renormalization of effective
interactions  replacing  the bare matrix $U_{12,34}=$ $%
\left\langle 12\left| v\right| 34\right\rangle $ in Eqs. (\ref{sig2},\ref{sigph})
with the corresponding scattering $T-$matrix averaged over spins

\[
\overline{U}=\left[ 1+U*\Pi _0(i\omega =0)\right] ^{-1}*U 
\]
where $\Pi _0(i\omega )$ is the Fourier transform of

\[
\Pi _{12,34}^0\left( \tau \right) =\frac 14\sum_{\sigma \sigma ^{\prime
}}G_{13}^\sigma \left( \tau \right) G_{24}^{\sigma ^{\prime }}\left( \tau
\right) 
\]
In the case of the single-band Hubbard model this approximation was found to be
very reliable \cite{fleck}. The effects of the spin- dependence of the effective
interaction matrix $\overline{U}$ will be considered elsewhere \cite{SPT}.

\section{Computational results}

We have started from the spin-polarized LSDA band structure of ferromagnetic
iron within the TB-LMTO method \cite{OKA} in the minimal $s,p,d$ basis set and
used numerical orthogonalization to find the $H_t$ part of our starting
Hamiltonian. We take into account Coulomb interactions only between $d$%
-states. The correct parameterization of the $H_{U\text{ }}$part is indeed a serious problem.
For example, first-principle estimations of average Coulomb
interactions (U) \cite{U,steiner} lead to unreasonably large value of order of
5-6 eV in comparison with experimental values of the U-parameter
in the range of 1-2 eV for iron\cite{steiner}. 
Semiempirical analysis of the appropriate  interaction value 
\cite{oles} gives $U\simeq 2.3$ eV. The difficulties with choosing  the
correct value of $U$ are connected with complicated screening problems,
definitions of orthogonal orbitals in the crystal, and contributions of the
intersite interactions. In the quasiatomic (spherical) approximation the
full $U$-matrix for the $d-$shell is determined by the three parameters $U, J$ and $%
\delta J$ \cite{lda++}. Note that the value of intra-atomic (Hund) exchange
interaction $J$ is not sensitive to the screening  and  approximately
equals to  0.9 eV in different estimations\cite{U}. We
use the simplest way of estimating  $\delta J$ keeping the ratio $\delta
J/J$  equal to its atomic value \cite{solj}. For  the most
important parameter $U$, which defines the bare vertex
matrix (Eq. (\ref{vertex})),
we use the value $U=2.3$ eV for  most of our calculations and discuss the
dependences of the density of states (DOS) as functions of $U$. To calculate
the spectral functions

\[
A_\sigma \left( {\mathbf k},E\right) =-\frac 1\pi Tr_LG_\sigma \left({\mathbf%
k},E+i0\right) 
\]
and DOS as their sum over the  Brillouin zone we first made analytical
continuation for the matrix self-energy from Matsubara frequencies to the real
axis using the Pade approximation \cite{pade}, and then numerically inverted 
 the Green-function matrix as in Eq. (\ref{dyson})
for each $\mathbf{k}$-point.
In the self-consistent solution of the FLEX equations we used 1024 Matsubara
frequencies and the FFT-scheme with the energy cut-off at 100 eV. The sum over
irreducible Brillouin zone have been made with 72 k-points for SCF-iterations
and with 1661 k-points for the final total density of states.

First we analyze the  $U$-dependence of electronic structure (Fig.1). Keeping $%
J, \, \delta J$ fixed as described above we vary the average $U$ parameter in the
range from 2 to 6 eV. For computational simplicity a relatively high temperature
value $T=1500$ K was used. Note that the temperature in this approach
is defined in terms of Matsubara frequencies.
In contrast to the standard LDA calculations at the finite temperatures \cite
{TLDA} we take into account the temperature dependence of the  Bose degrees of
freedoms through the p-h susceptibilities. It is known \cite{moriya} that this source of
temperature dependence is the most important for  itinerant electron magnets.
Although  T is above T$_C^{\exp }$ (where T$_C^{\exp }=1043$
K is the experimental value of the Curie temperature of iron) all DOS ($%
N_{\uparrow }\left( E\right) ,N_{\downarrow }\left( E\right) $) curves in
the Fig.1 show the spin splitting. In principle, one could calculate T$_C$
using the temperature dependence of the uniform spin susceptibility \cite{dinf}
but we have not done it yet. Nevertheless, we believe that this splitting in
 DMFT is a manifestation of the existence of local magnetic
moment above $T_C$ and is not connected with long-range magnetic order. This
spin splitting is a characteristic feature of local approximations and
reflects the Hubbard splitting in the one-band model \cite{fleck}. The local
spin splitting above Curie temperature in iron is definitely observed in 
many experiments, e.g. optics, photoemission etc. (see, e.g. \cite
{moriya,IKT} and references therein).

First of all, we see that the value of the total magnetic moment is weakly
dependent on $U$ and is of order of $2\mu _B$ for this temperature. The
positions of the main peaks with respect to the Fermi energy roughly coincides with
those in LSDA up to $U=2$ eV. Starting from $U=3$~eV satellites at $E\simeq
-5 $ eV appears as well as additional many-body structure at $E\simeq 4$~eV.
 Note that a weak satellite-like feature at $E\simeq -5$ eV
was observed experimentally in \cite{satyes} although it was not found in \cite{kirby}%
. For $U\simeq 1$ eV which is considered to be ``experimental''
value for iron \cite{steiner} there are no any noticeable manifestations of
this satellite. The boundary values of $U$ of order of 2 eV
with weak shoulder probably corresponds to the experimental situation in
the best way. For $U\simeq 5-6$ eV an essential part of the spectral density
related to the many-particle peaks corresponding to the upper and lower Hubbard
bands, which is  unrealistic for such moderately correlated
substances as iron. At $U\simeq 6$ eV the empty quasiparticle minority-spin peak
goes below the Fermi level which decreases the magnetic
moment.

The depolarization of states near the Fermi level is
another important correlation effect. 
The decrease of the ratio $P=\left[ N_{\uparrow }\left(
E_F\right) -N_{\downarrow }\left( E_F\right) \right] /\left[ N_{\uparrow
}\left( E_F\right) +N_{\downarrow }\left( E_F\right) \right] $  is a
typical sign of spin-polaron effects \cite{IKT,uspekhi}.
In our approach this effects  are
taken into account through the $W_{\uparrow \downarrow }^{(ph)}$
terms in the effective spin-polarized LDA++ potential.

The energy dependence of self-energy in Fig.2 shows characteristic features
of moderately correlated systems. At low energies $\left| E\right| <1$ eV we
see a typical Fermi-liquid behavior $Im\Sigma \left( E\right) \sim -E^2,$ $%
\partial Re\Sigma \left( E\right) /\partial E<0.$ At the same time, for the
states beyond this interval within the $d$-bands the damping is rather large
(of the order of 1 eV) so these states corresponds to ill-defined
quasiparticles, especially for occupied states. This is probably one of the
most important conclusions of our calculations. Qualitatively it was already
pointed out in Ref. \cite{treglia} on the basis of a model second-order
perturbation theory calculations. We have shown that this is the case
of realistic quasiparticle structure of iron with the reasonable value of Coulomb
interaction parameter.

Due to noticeable broadening of quasiparticle states the description of the
computational results in terms of effective band structure (determined, for
example, from the maximum of spectral density) would be incomplete. We
present on the Fig.3 the \textit{full} spectral density $A_\sigma \left( 
{\mathbf k}, E\right) $ including both coherent and incoherent parts as a
function of $\mathbf{k}$ and $E$. We see that in general the maxima of the
spectral density (dark regions)
coincide with the experimentally obtained band structure.
However, for occupied majority spin states at about -3 eV the distribution
of the spectral density is rather broad and the  description 
of this states in terms
of the quasiparticle dispersion is problematic. This conclusion is in
complete quantitative agreement with raw experimental data on angle-resolved
spin-polarized photoemission \cite{kisker} with the broad non-dispersive
second peak in the spin-up spectral function around -3 eV.

\section{Applications to spin-polarized thermoemission}

One of the most unexpected  results concerning electronic
structure of iron was obtained by spin-polarized
thermoemission for cesiated iron\cite{thermoem}.
In this case the thermal current is determined 
 by the states with the energy $W=1.37$ eV above Fermi level which are in
the region of quasiparticle DOS peak for minority spin. One could expect a
strong negative spin polarization of the current 
(polarization ratio P as estimated from LSDA DOS is about -85\% ).
 More accurate estimation which takes into account group velocities 
\cite{monnier} results in P= -34\% for the polarization ratio.
Experimentally it found to be zero within the experimental error.

To clarify the situation, we considered this effect on the basis of our LDA++
calculations. The current through the surface $x=0$ in the
spectral representation is

\begin{equation}
j_x=\sum_{\mathbf{k}}Tr_{L\sigma }\left[ \frac{\partial t\left( \mathbf{k}%
\right) }{\partial k_x}\left\langle c_{\mathbf{k}}^{+}c_{\mathbf{k}%
}\right\rangle \right] =\sum_{\mathbf{k}}Tr_{L\sigma }\left\{ \frac{\partial
t\left( \mathbf{k}\right) }{\partial k_x}\left[ \int\limits_{-\infty
}^\infty dEf(E)A({\mathbf k}, E)\right] \right\}  \label{therm1}
\end{equation}
where $\partial t\left( \mathbf{k}\right) /\partial \mathbf{k}=\mathbf{V}_{%
\mathbf{k}}$ is the group velocity operator (the matrix in orbital
indices), $f(E)$ is the Fermi distribution function. Taking into account
only electrons moving towards the surface $\left( k_x>0\right) $ 
and with the  energy (E)
above the barrier (W) the thermoemission current (cf.  Ref\cite
{monnier} for the case of noninteracting electrons)
could be expressed as:

\begin{equation}
j_x^T=\sum_{{\mathbf k} (k_x>0)} \int\limits_W^\infty dEf(E)Tr_{L\sigma }\left\{
\left| \frac{\partial t\left( \mathbf{k}\right) }{\partial k_x}\right| A(%
{\mathbf k}, E)\right\}   \label{therm2}
\end{equation}
Taking into account that $T<<W$ and averaging over the
surface orientations (experiments \cite{thermoem} were carried out for
 polycrystalline
samples), we found the following approximate formula for the polarization

\begin{eqnarray}
P &=&\frac{I_{\uparrow }-I_{\downarrow }}{I_{\uparrow }+I_{\downarrow }},
\label{therm3} \\
I_\sigma  &=&\sum\limits_{i=x,y,z}\sum_{\mathbf{k}}Tr_L\left\{ \left| \frac{%
\partial t\left( \mathbf{k}\right) }{\partial k_i}\right| A_\sigma ({\mathbf%
k}, W)\right\}   \nonumber
\end{eqnarray}
This was calculated using numerical differentiation of 
$t\left( \mathbf{k}\right) $ matrix and  summing up over 1661 $\mathbf{k}$
points in the irreducible part of the Brillouin zone. We obtained $P=$-12\% 
which is to compare with the value of $-34\%$ from LSDA calculations \cite{monnier}.
The decrease of P is not a pure effect of damping of the
quasiparticle states, but is the result of rather complicated 
cancellations of $s,p,d$ -electron contributions. Therefore, one 
may conclude that there is no drastic discrepancy between experimental
results \cite{thermoem} and theoretical description of the electronic
structure of iron, inspite of the 
approximate character of our treatment of the thermoemission problem.
For  more accurate description one needs to consider 
, for example, an influence of the cesium layer and surface effects on the
electronic structure of iron according to the experimental conditions.

\section{Conclusions}

We have proposed a general scheme for investigation of the correlation effects
in the quasiparticle band structure calculations for itinerant-electron
magnets. This approach is based on the combination of the dynamical mean-field theory and
the fluctuating exchange approximation. 
Application of LDA++ method  gives an adequate
description of the quasiparticle electronic structure for ferromagnetic iron.
The main
correlation effects in the electron energy spectrum are strong damping of
the occupied states below 1 eV from the Fermi level $E_F$ and essential
depolarization of the states in the vicinity of $E_F$. We obtained a
reasonable agreement with different experimental spectral data 
(spin-polarized photo- and thermoemission). The method is rather
universal and can be applied for other magnetic systems, both ferro- and
antiferromagnets.

\section{Acknowledgments}

Part of this work was carried out during the visit of one of the author (MIK)
to Max Planck Institute of Physics for Complex Systems (Dresden). The work
was also partially supported by Russian Basic Research Foundation under
grant 96-02-16000. We are grateful to O.K. Andersen, C. Carbone, P. Fulde,
O. Gunnarsson, G. Kotliar and A. Georges for helpful discussions.


\begin{figure}
\centerline{\epsfig{file=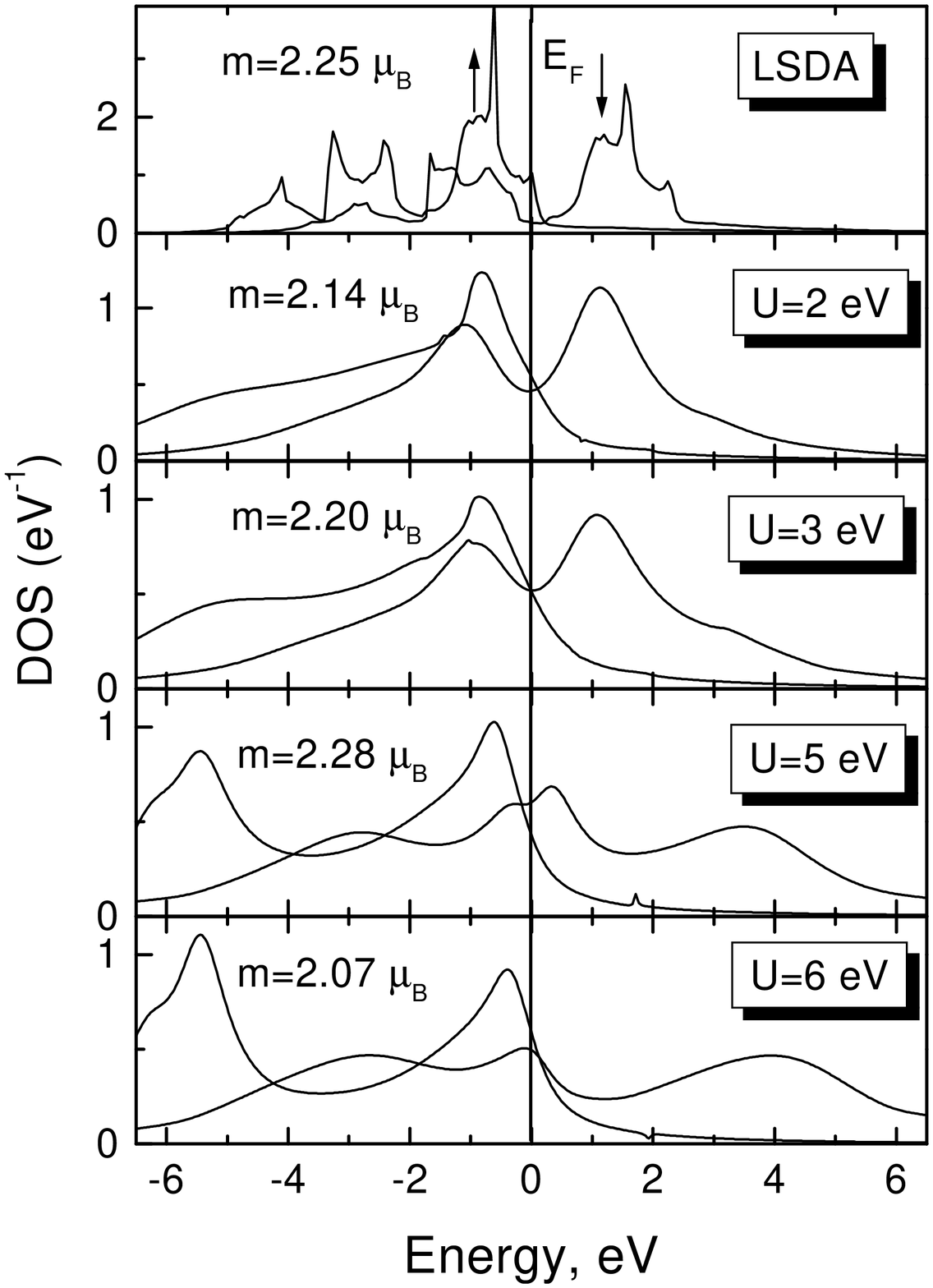,width=12cm}}
\vskip  1cm
\caption{
Density of d-states and magnetic moments for ferromagnetic iron in the LSDA andthe LDA++ calculations
for different average Coulomb interactions with J=0.9 eV
and temperature T=1500 K.  
}
\label{DOS}
\end{figure}

\begin{figure}
\centerline{\epsfig{file=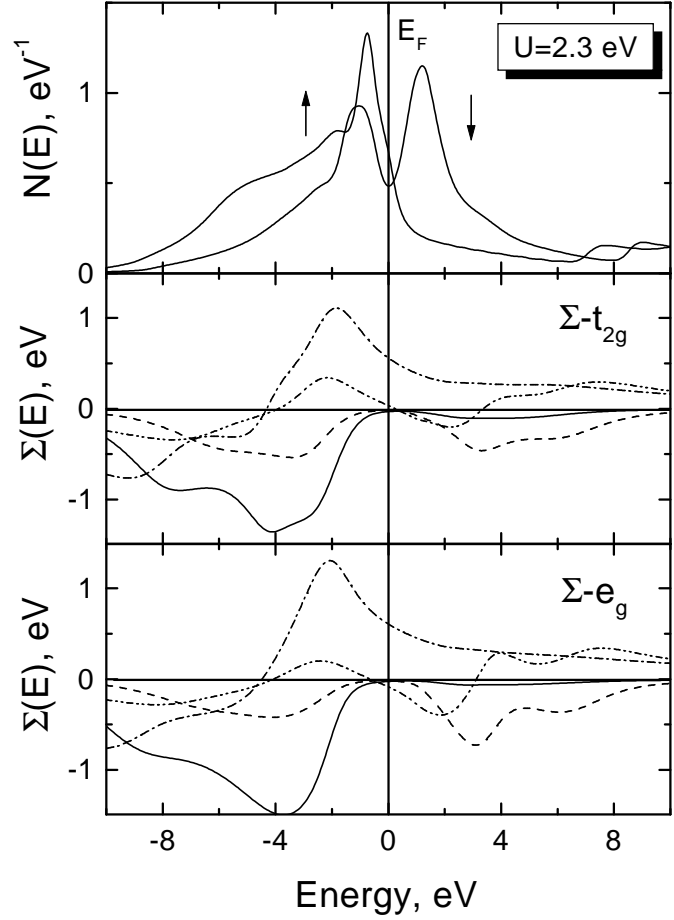,width=10cm}}
\vskip  1cm
\caption{
Total density of states and d-part of self-energy for ferromagnetic iron 
with  U=2.3 eV and J=0.9 eV for the
temperature T=750 K. Two different self-energies for t$_{2g}$ and
e$_g$ d-states in the cubic crystal field symmetry are presented 
and  four different lines corresponds to imaginary part spin-up (full line)
and spin-down (dashed line) as well as real part spin-up (dashed-dot line)
and spin-down (dashed-double-dot line).
}
\label{SIG}
\end{figure}

\begin{figure}
\begin{minipage}{7.5cm}
\vskip  0cm
  \epsfig{file=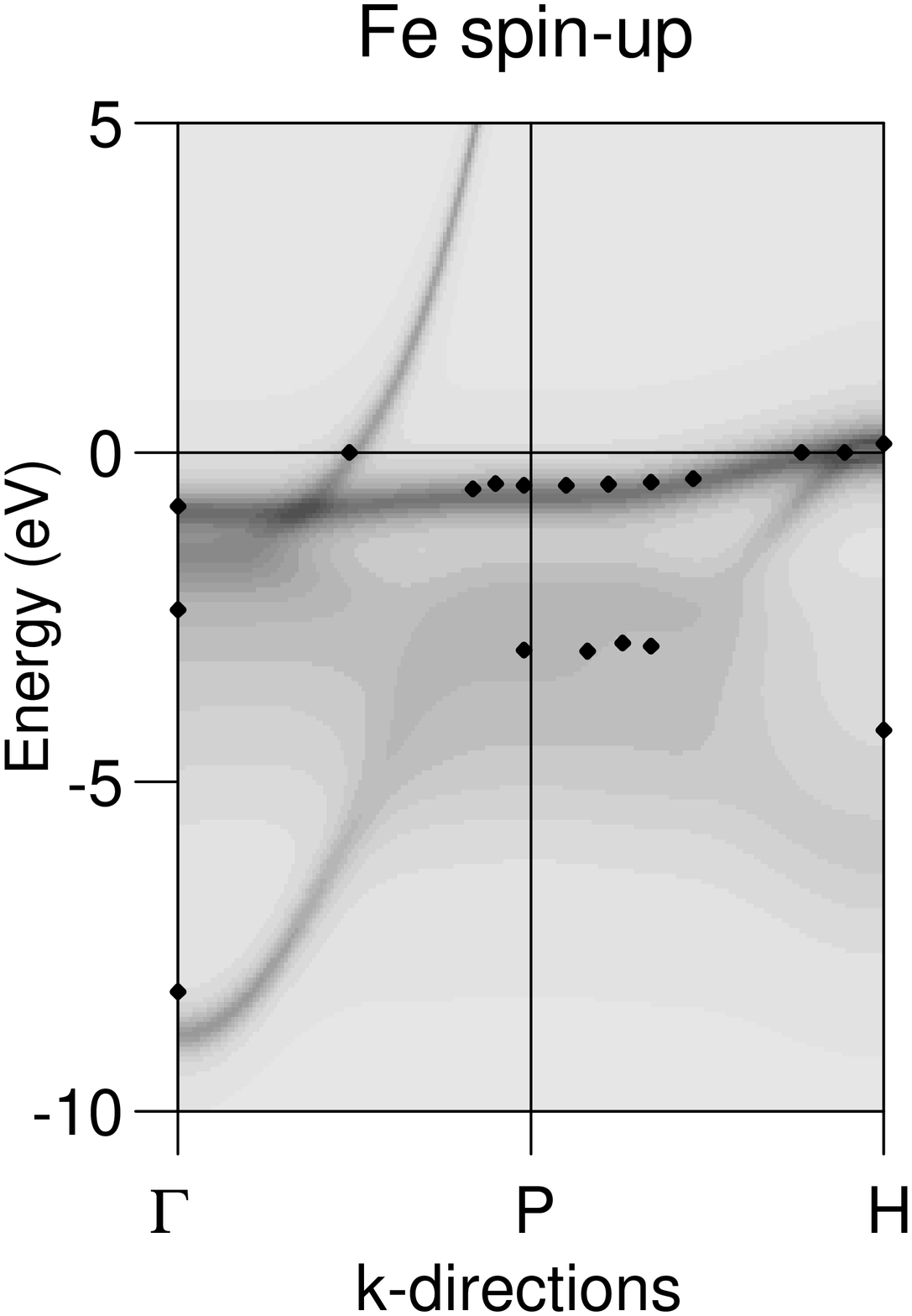,width=7.5cm}
\end{minipage}
\hspace{0.1cm}
\begin{minipage}{7.5cm}
  \epsfig{file=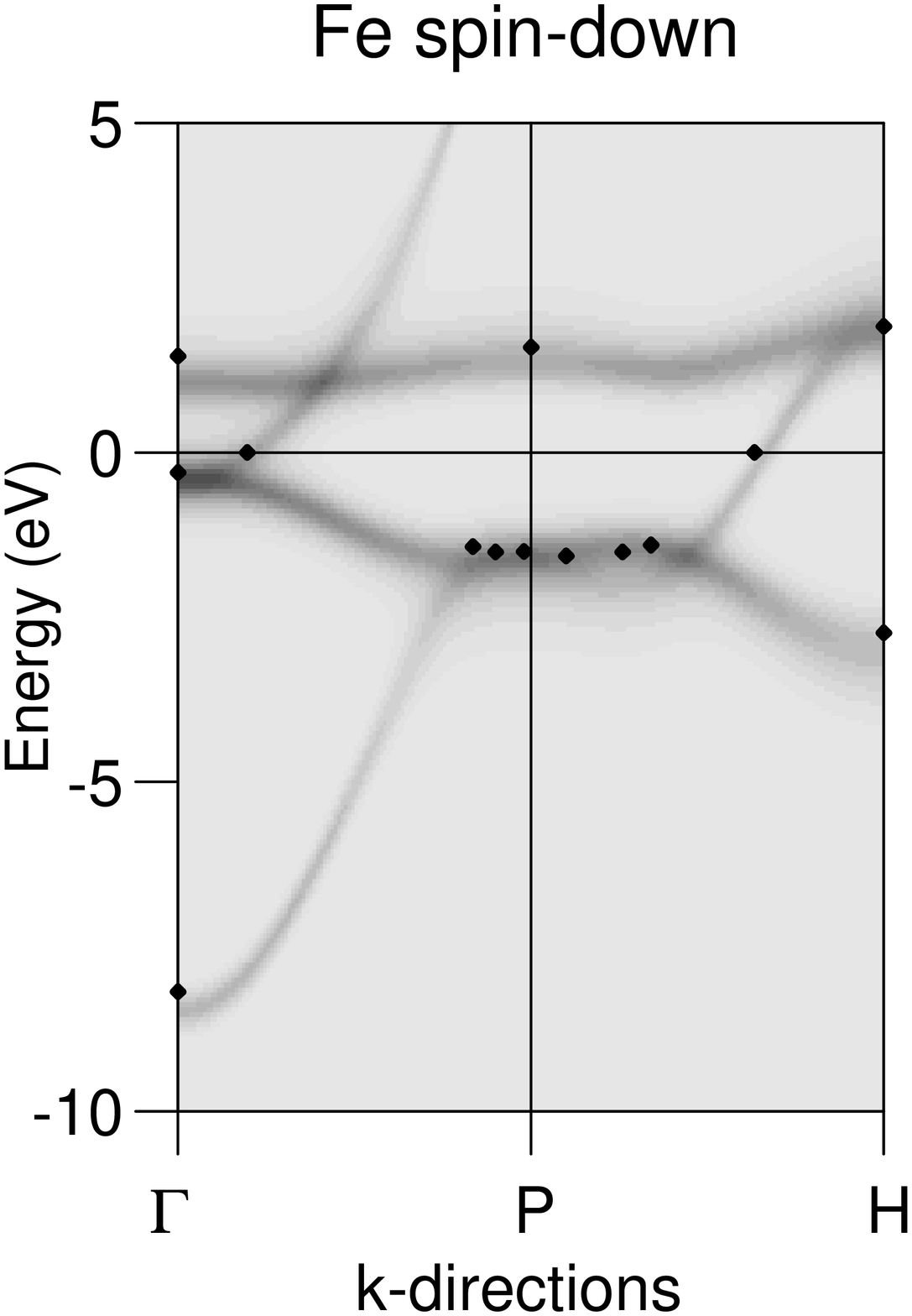,width=7.5cm}
\end{minipage} \\
\vskip 0.3cm
\caption{
Spectral function of ferromagnetic iron for spin-up (a) and spin-down (b)
and the two k-directions in the Brillouin zone compare with the experimental
angle resolved photoemission and de Haas - van Alphen (at the E$_F$=0)
points (from Ref. 3 ).
}
\label{Ak2}
\end{figure}

\end{document}